\documentclass[prd,preprint]{revtex4}
\usepackage{graphicx}
\usepackage{amssymb}

\begin{document}

\title{Remarks on the Cosmic Density of Degenerate Neutrinos}
\author{Kazuhide Ichikawa}
\author{M. Kawasaki}
\affiliation{Research Center for the Early Universe, University 
  of Tokyo, Bunkyo-ku, Tokyo 113-0033, Japan}
\date{\today}

\begin{abstract}
We re-investigate the evolution of the strongly degenerate neutrinos in the early universe. With the larger degeneracy, the neutrino number freezes at higher temperatures because the neutrino annihilation rate decreases. We consider  very large degeneracy so large that the neutrino number freezes before events in which the particle degrees of freedom in the universe decrease (e.g. the muon annihilation and the quark-hadron phase transition). In such a case, the degeneracy by the time of nucleosynthesis becomes smaller than the initial degeneracy. We calculate how much it decreases from the initial value on the basis of the conservation of the neutrino number and the total entropy. We found a large drop in the degeneracy but it is not large enough to affect the current constraints on the neutrino degeneracy from BBN and CMBR.
\end{abstract}

\maketitle

\section{introduction}
It is well known that the evolution of the universe deeply depends on the properties of neutrinos (see Ref.~\cite{Dolgov2002} for review). In the standard cosmology, we assume three types of massless neutrinos with the same number of particles and antiparticles. In this paper, we deviate from the last assumption and consider cosmological effects caused by neutrino-antineutrino number asymmetry. We use a terminology "degeneracy" for this asymmetry. We also assume, just for simplicity of the description, the neutrino number is larger than the antineutrino number ({\it i.e.~}the neutrino has positive chemical potential). When considering the opposite case, we need to just exchange the role of neutrinos and antineutrinos.  

Since degeneracy increases the sum of the energy density of neutrinos and antineutrinos, it significantly affects standard predictions such as the big bang nucleosynthesis and the cosmic microwave background (in addition, electron-type neutrinos destroy neutrons so the nucleosynthesis depends strongly on their number density). Naively, these effects monotonously increase with the degree of degeneracy so the observations can put upper bounds on the degeneracy in each generation of neutrinos. 

However, things are not as simple as this because the neutrino number freeze-out temperature also increases with the degeneracy (the rise in the freeze-out temperature is caused by the scarcity of the antineutrinos, the annihilation partners of the neutrinos). The complication occurs when the freeze-out takes place before the muon-antimuon annihilation ends. In this case, when muons and antimuons annihilate, already frozen neutrino number can not change but the entropy is transfered to the neutrinos through the elastic scattering with electrons and positrons so the neutrinos  keep the same temperature with the rest of the cosmic plasma. In order to conserve the number while the temperature changing faster than the inverse of the scale factor, the neutrino degeneracy parameter, chemical potential divided by temperature, decreases. As a result, by the onset of nucleosynthesis, the neutrino energy density becomes much lower than the value calculated using the initial degeneracy. Therefore, there is a logical possibility that very large initial degeneracy cannot be ruled out by the observations.

This possibility is pursued in Ref.~\cite{Kang1992} but with an incorrect picture on the "neutrino decoupling" when very large degeneracy exists (the same lines of argument are still found in a few papers \cite{Orito2000} \cite{Orito2002} \cite{etc}). They have regarded the neutrino number freeze-out and its kinematical decoupling from the rest of the cosmic plasma occur simultaneously. The correct picture is the one as described in the previous paragraph, the kinetic equilibrium holds well after the chemical equilibrium ceases to hold. The reason is that there is small number of annihilation partners, antineutrinos, but elastic scattering partners, electrons and positrons, are abundant. This is pointed out in Ref.~\cite{Dolgov2002}, but they have not calculated the relation between the initial degeneracy and the final degeneracy after the entropy producing events. Obtaining this relation is the main purpose of this paper. 

In the next section, we calculate the neutrino number freeze-out temperature and show it increases exponentially with the initial degeneracy parameter. We also justify the neutrinos and anti-neutrinos are in kinetic equilibrium during the annihilation by demonstrating their elastic scattering rate is sufficiently large. In Sec.~\ref{sec:change}, for a certain range of initial degeneracy parameter, we calculate their final value after the entropy producing events using the neutrino number conservation and the total entropy conservation. This is our main result. In Sec.~\ref{sec:conclusion}, we summarize the results and discuss the current constraints on the neutrino degeneracy are not affected.

\section{neutrino number freeze-out} \label{sec:freezeout}

We calculate the neutrino number freeze-out temperature $T_f$ by $\Gamma_A(T_f)=H(T_f)$, where $\Gamma_A$ is the rate of the change in the neutrino number density $n_{\nu}$ through the neutrino-antineutrino annihilation processes and $H$ is the cosmic expansion rate. 

In order to calculate $\Gamma_A$, we have to sum all the annihilation process rate. Assuming one type of the neutrino $\nu_{l}$ has degeneracy, we consider the annihilation to some fermion-antifermion pairs, $\nu_{l}(p_1) + \bar{\nu}_{l}(p_2) \rightarrow F(p_3) + \bar{F}(p_4)$ where the variables in the brackets denote the four-momentum of each particle in the comoving frame. For $F$, there are electrons and two non-degenerate types of neutrinos. In addition, we consider the annihilation to muon-anitimuon pairs when $T>m_{\mu}/3$ and to quark-antiquark pairs when $T>T_{\rm QCD}$ where we assume the quark-hadron phase transition to occur instantaneously at $T_{\rm QCD}=200$ MeV and the quark-gluon phase contains u, d and s quarks. The contribution of this process to the $\Gamma_A$ is,
\begin{eqnarray}
\Gamma_{\nu_l\bar{\nu_l}\rightarrow F\bar{F}}&=&-\left( \frac{\dot{n}_{\nu_l}}{n_{\nu_l}} \right)_{\nu\bar{\nu}\rightarrow F\bar{F}} \nonumber \\
&=&-\frac{1}{n_{\nu_l}} \int \frac{d^3p_1}{(2\pi)^3} \frac{d^3p_2}{(2\pi)^3} \frac{d^3p_3}{(2\pi)^3} \frac{d^3p_4}{(2\pi)^3} 
|{\cal M}(\nu_l + \bar{\nu_l} \rightarrow  F + \bar{F})|^2 \nonumber \\
& &\times(2\pi)^4 \delta^{(4)}(p_1+p_2-p_3-p_4)f_{\nu_l}(E_1) f_{\bar{\nu_l}}(E_2) [1-f_{F}(E_3)] [1-f_{\bar{F}}(E_4)]. \label{eq:rateann}
\end{eqnarray}
$F$'s are well-approximated to be massless so the square of the invariant matrix element $|{\cal M}(\nu_l + \bar{\nu_l} \rightarrow  F + \bar{F})|^2$ can be written in the form $32G_F [b(p_1 {\cdot} p_3)^2 + c(p_1 {\cdot} p_4)^2]$ where $G_F=(292.80 {\rm GeV})^{-2}$ is the Fermi coupling constant. For $F\neq l$, only the neutral current contributes so that $b=(C_V^F-C_A^F)^2$ and $c=(C_V^F+C_A^F)^2$ and for $F=l$, the charged current contributes in addition so that $b=(C_V^l-C_A^l)^2$ and $c=(C_V^l+C_A^l+2)^2$. The vector and axial-vector couplings $(C_V^F,C_A^F)$ are $(1/2,1/2)$ for $F=\nu$, $(-1/2,-1/2+2\sin^2\theta)$ for $F=e$ and $\mu$, $(1/2,1/2-(4/3)\sin^2\theta)$ for $F=u$, and $(-1/2,-1/2-(2/3)\sin^2\theta)$ for $F=d$ and $s$ where the weak-mixing angle $\sin^2\theta=0.231$. $f$'s are the distribution functions of the particle species on the subscript and we use the equilibrium form, $f_{F,\bar{F}}=1/[\exp(E/T)+1]$ and $f_{\nu_l,\bar{\nu_l}}=1/[\exp((E\pm\mu)/T)+1]$ where $\mu$ is the chemical potential and the signs are $-$ for $\nu_l$ and $+$ for $\bar{\nu_l}$. Here, we assume $\mu=\mu_{\nu_l}=-\mu_{\bar{\nu_l}}$, expecting the chemical equilibrium holds initially at very high temperature. Some of the integration can be performed analytically and we obtain 
\begin{eqnarray}
\dot{n}_{\nu_l}|_{\nu_l\bar{\nu_l}\rightarrow F\bar{F}} &=&T^8\frac{G_F^2}{64\pi^5} \int_0^{\infty} dx  \int_0^{\infty} dy \int_{-1}^{1} dz \int_{f_-}^{f_+} dt\ f_{\nu_l}(x) f_{\bar{\nu_l}}(y) x^3 y^3(1-z)^2\delta^{-5} \nonumber \\
& & \times \Big[ (b+c)\left\{ 3\delta^{4}+3(x-y)^2(2t-x-y)^2-(x-y)^2\delta^{2}-(2t-x-y)^2\delta^{2} \right\} \nonumber \\
& &-4(b-c)(x-y)(2t-x-y)\delta^{2}  \Big] [1-f_{F}(t)][1-f_{\bar{F}}(x+y-t)] \nonumber \\
&\equiv&T^8 G_F^2 L(\xi) \label{eq:dnnudt}
\end{eqnarray} 
where $x=E_1/T$, $y=E_2/T$, $z=\cos\theta$, $t=E_3/T$, $\delta(x,y,z)=(x^2+y^2+2xyz)^{1/2}$ and $f_{\pm}=(x+y\pm\delta)/2$. With these variables, the distribution functions are $f_{\nu_l,\bar{\nu_l}}(x)=1/[\exp(x\pm\xi)+1]$ and $f_{F,\bar{F}}(x)=1/[\exp(x)+1]$, where we define the degeneracy parameter $\xi\equiv \mu/T$. Finally, the degenerate neutrino number density is 
\begin{equation}
n_{\nu_l}(\xi)=\frac{T^3}{2\pi^2}\int\frac{x^2}{e^{x-\xi}+1} dx \equiv M(\xi)T^3. \label{eq:nnu}
\end{equation}

The cosmic expansion $H$ is determined by the total energy of the universe by the Einstein equation,
\begin{equation}
H=\sqrt{\frac{8\pi\rho_{tot}}{3M_P^2}}=\sqrt{\frac{4\pi^3 g_{\ast}(\xi)}{45}}\frac{T^2}{M_{pl}}, \label{eq:hubble}
\end{equation}
where $M_P=1.2\times10^{19}$ GeV is the Planck energy and we have introduced $g_{\ast}$ to express the total energy density as $\rho_{tot}=g_{\ast}(\xi)(\pi^2/30)T^4$. On calculating the total energy, we do not have to worry about temperature dependence of $g_{\ast}$ because it turns out that when the freeze-out temperature becomes so high as muons to contribute as relativistic degree of freedom, $\xi$ should be large making the muon (and the other particle species other than the degenerate neutrino) energy density negligible compared to that of the degenerate neutrino. The same is true for pions and the quark-gluon plasma. Therefore we include the contribution from photons, electrons, positrons, two types of non-degenerate (anti)neutrinos and the degenerate (anti)neutrinos, 
\begin{eqnarray}
g_{\ast}(\xi)=9+ \frac{15}{\pi^4}\int \left( \frac{x^3}{e^{x-\xi}+1} + \frac{x^3}{e^{x+\xi}+1} \right)dx. \label{eq:gstar}
\end{eqnarray}
This gives usual value $g_{\ast}(0)=43/4$.

\begin{figure}
\includegraphics{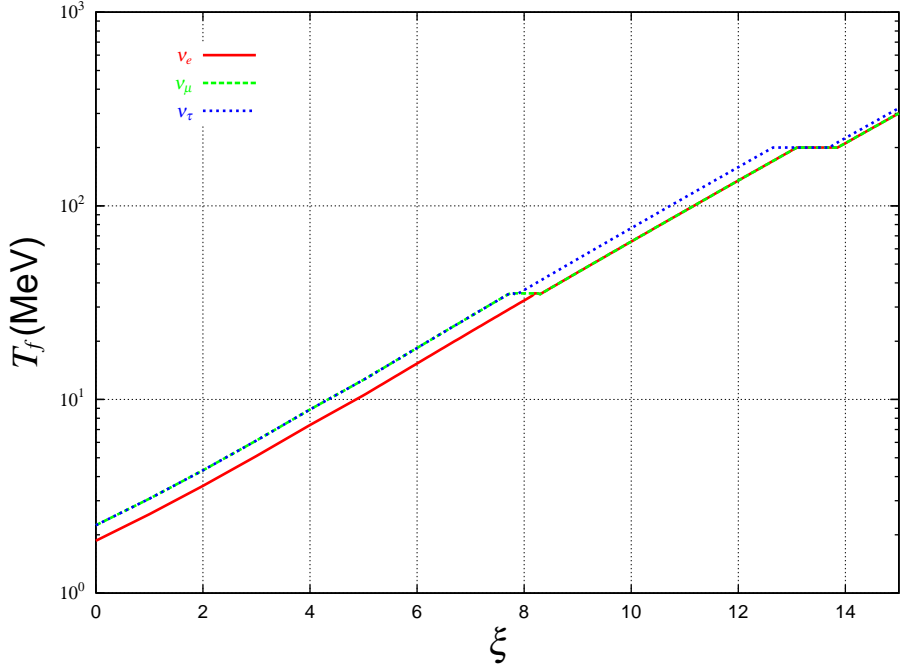}
\caption{The relation between the neutrino degeneracy and its number freeze-out temperatures. The solid, dotted and small-dotted lines respectively show  when $\nu_e$, $\nu_{\mu}$ and $\nu_{\tau}$ has the degeneracy. They basically increase exponentially with the degeneracy. At $T\sim m_{\mu}/3$ and $T_{\rm QCD}$, according to the variation in the relativistic freedom of the fermions, muons and quarks, to which neutrinos can annihilate, the change in the freeze-out temperature with degeneracy is retarded.}
\label{fig:Tf}
\end{figure}

Combining Eqs.~(\ref{eq:rateann}) $\sim$ (\ref{eq:gstar}) gives freeze-out temperatures of the neutrino number as a function of $\xi$
\begin{eqnarray}
T_f(\xi)=0.999756\ g_{\ast}(\xi)^{1/6} \left[\frac{M(\xi)}{L(\xi)}\right]^{1/3} \quad{\rm MeV}. \label{eq:Tfexact}
\end{eqnarray}
The results are shown in Fig.~\ref{fig:Tf}. We find that $T_f$ increases exponentially with $\xi$. This is expected because the number density of the antineutrinos, the annihilation partners of neutrinos, is exponentially suppressed due to the degeneracy. To see this more explicitly and for the later convenience, we here derive the approximate expression for $T_f$ when $\xi$ is large. First, Eq.~(\ref{eq:dnnudt}) is simplified by setting $f_{\bar{\nu_l}}(y) \approx \exp(-y) \exp(-\xi)$ and neglecting $e^{\pm}$ Pauli blocking factors. Using the integration approximation formula for a function satisfying $\varphi(0)=0$, 
\begin{equation}
\int_0^{\infty} \varphi^{\prime}(x)f_{\nu}(x)dx=\int_0^{\infty} \frac{\varphi^{\prime}(x)dx}{\exp(x-\xi)+1}=\varphi(\xi)+\frac{\pi^2}{6}\varphi^{\prime\prime}(\xi)+\frac{7\pi^4}{360}\varphi^{(4)}(\xi)+\cdots ,
\end{equation}
we obtain
\begin{eqnarray}
L(\xi) &\approx& \frac{2}{3\pi^5}(b+c)\ \exp(-\xi) \left[ \frac{\xi^4}{4}+\frac{\pi^2 \xi^2}{2}+\frac{7\pi^4}{60} \right], \\
M(\xi) &\approx& \frac{\xi}{6} \left[ \frac{\xi^2}{\pi^2} + 1 \right].
\end{eqnarray}
For $g_{\ast}(\xi)$, we neglect the exponentially suppressed contribution from antineutrinos and obtain 
\begin{eqnarray}
g_{\ast}(\xi) \approx  \frac{43}{4} + \frac{15}{4} \left[ \left(\frac{\xi}{\pi}\right)^4+2\left(\frac{\xi}{\pi}\right)^2 \right] .
\end{eqnarray}
We have the expression for $T_f$ by putting these approximations into Eq.~(\ref{eq:Tfexact}) and it reproduces closely the numerical result plotted in Fig.~\ref{fig:Tf} when $\xi \gtrsim 5$. 

Now we explicitly write down the expression for $T_f$ taking the leading order in $\xi$ to compare with that of Refs.~\cite{Dolgov2002}, \cite{Kang1992} and \cite{Freese1983}. Our analysis shows
\begin{eqnarray}
T_f \approx 1.37408 \left(\frac{b+c}{2.3432}\right)^{-\frac{1}{3}} \xi^{\frac{1}{3}} \exp(\xi/3)\ {\rm MeV}. 
\end{eqnarray}
This and theirs all agree with respect to $T_f \propto \exp(\xi/3)$ which originates from the exponentially suppressed degenerate $n_{\bar{\nu}}$. Our expression contains the factor $\xi^{1/3}$,  which originates from the degenerate $\nu$ distribution affecting both of $n_{\nu}$ and $H$, but it seems not to appear in the others. However, the expression of Ref.~\cite{Dolgov2002} agrees with ours if thermally averaged momentum divided by $T$, $\langle p/T \rangle$, is calculated using the degenerate $\nu$ distribution so that $\langle p/T \rangle \approx (3/4)\xi$ (they use $\langle p/T \rangle \approx 3$ which is true when the degeneracy is small) and $g_{\ast}(\xi)$ is approximated as $\propto \xi^4$. As for Ref.~\cite{Freese1983}, $T_f \propto \xi^{-2/3}$ but this is thought to originate from the incorrect division by $n_e$ when deriving $\Gamma_A$. This does not make difference when calculating $T_f$ with no degeneracy but is not appropriate when strong degeneracy exists. Ref.~\cite{Kang1992} has not found such a factor of power of $\xi$ but this is thought to originate from calculating the annihilation rate in the center-of-mass frame. In addition, they also calculate $\dot{n_e}/n_e$ so,  as Ref.~\cite{Orito2000} has pointed out, their result has to be corrected anyway.

Until now, we have not cared about the possibility of the neutrino spectral distortion and have used thermal distributions to calculate various quantities. We now justify this procedure. Since neutrinos with larger energies have more probability to annihilate (from Eq.~(\ref{eq:dnnudt}), we see $\dot{f_{\nu}} / f_{\nu} \propto E_1$), they seems to freeze-out later and the neutrino spectrum would distort very much. But as we discuss next, the neutrino-electron elastic scattering occur sufficiently so the thermal distribution of neutrinos is preserved at the time of annihilation. Since the presence of neutrino large degeneracy ensures that the neutrino-antineutrino annihilation process occurs with their distribution preserving equilibrium form, it is considered that the estimation of neutrino number freeze-out temperature from $\Gamma_A=H$ turns out to be good. 

We investigate how these degeneracy parameters undergo a change in the next section. Before we proceed, we estimate the temperature at which the neutrinos decouple kinematically from $e^{\pm}$. This temperature is expected to be much lower than the number freeze-out temperature because the partners of the elastic processes, electrons and positrons are abundant regardless of the neutrino degeneracy. So they are coupled until relatively low temperature and have the same temperature with the others even after their number has frozen out at higher temperature. This decoupling temperature is well estimated from the rate for the elastic scattering processes because the energy transfer between $e^{\pm}$ and $\nu$ occurs efficiently in this case. The energy transfer efficiency is estimated as follows. The center-of-mass frame differencial cross section for the elastic processes is
\begin{equation}
\left( \frac{d\sigma_{CM}}{d\cos\theta} \right)_{\nu + e^{-} \rightarrow \nu + e^{-}} +
\left( \frac{d\sigma_{CM}}{d\cos\theta} \right)_{\nu + e^{+} \rightarrow \nu + e^{+}}= \frac{s}{16\pi}G_F^2(C_V^2+C_A^2) \left\{ 4+(1+\cos\theta)^2 \right\}, \label{eq:cselastic}
\end{equation}
where $s$ and $\theta$ are the total initial energy squared and the polar angle in the center-of-mass frame. The energy transfer from $e^{\pm}$ to $\nu$ with initial energy (in the laboratory frame) $E_{e^{\pm}}$ and $E_{\nu}$ is
\begin{equation}
E_{{\rm transfer}}=\frac{1}{2}(E_{e^{\pm}} - E_{\nu})(1-\cos\theta). \label{eq:etransfer}
\end{equation}
From (\ref{eq:cselastic}) and (\ref{eq:etransfer}), the expectation value of the energy transfer by one collision is computed as
\begin{equation}
\langle E_{{\rm transfer}} \rangle = \frac{\displaystyle{\int E_{{\rm transfer}} \frac{d\sigma}{d\cos\theta} d(\cos\theta)}}{\displaystyle{\int \frac{d\sigma}{d\cos\theta} d(\cos\theta)}} = \frac{7}{16}(E_{e^{\pm}} - E_{\nu}).
\end{equation}
Therefore, roughly speaking, $\nu$ becomes almost as energetic as $e^{\pm}$ after one collision.

\begin{figure}
\includegraphics{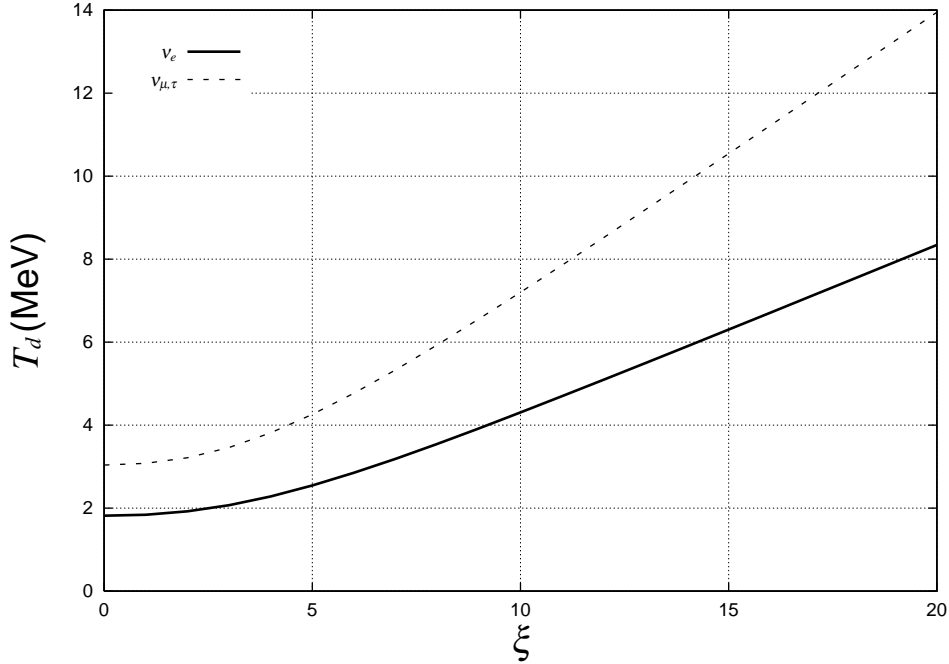}
\caption{The decoupling temperatures of the degenerate neutrinos. The neutrinos are in kinetic equilibrium with $e^{\pm}$ and $T_{\nu}$ follows $T_{e^{\pm}}$'s variation above these temperatures. Note that they increase with $\xi$ but do not exceed $m_{\mu}/3$.}
\label{fig:Td}
\end{figure}

Now, we compute the rate for the elastic scattering processes
\begin{equation}
\Gamma_{\nu,{\rm elastic}}=-\left[\left(\frac{dn_{\nu}}{dt}\right)_{\nu + e^- \rightarrow \nu + e^-}
                        + \left(\frac{dn_{\nu}}{dt}\right)_{\nu + e^+ \rightarrow \nu + e^+}\right] \Bigg/n_{\nu}
\end{equation}
and search a temperature $T_d$ at which it becomes equal to the expansion rate, just as we have done in the case of the annihilation process. The result is shown in Fig. \ref{fig:Td}. We see that $T_d$ increase with $\xi$ but rather slowly.  Therefore, as is the case of no degeneracy, the neutrinos are in kinetic equilibrium holding the same temperature as the rest of the plasma until at least the muon-antimuon annihilation ends.

\section{the change in chemical potential and energy density}\label{sec:change}

In this section, we study how the neutrino or its degeneracy parameter $\xi$ evolves after its number has frozen out. $\xi$ is conserved when the annihilation occurs frequently so that chemical equilibrium holds. Even after the annihilation practically ceases, $\xi$ does not vary as long as the temperature $T$ falls as $T \propto a^{-1}$ where $a$ is the cosmological scale factor. But this condition is not satisfied when, for example, the muon annihilation progresses because the temperature falls more slowly (recall from the end of the last section that the neutrino temperature evolves along with the $e^{\pm}$ temperature because they are in kinetic equilibrium). For the case when the freeze-out takes place before the muon annihilation ends, there is a period in which the neutrino number at the freeze-out temperature is conserved while the temperature falls slower than $a^{-1}$. This causes $\xi$ to decrease during the muon annihilation or other entropy producing processes. 

We can compute the value of $\xi$ after the muon annihilation on the basis of neutrino number and total entropy conservation after the freeze-out. We briefly explain why the latter holds. The second law of thermodynamics for the total gas in physical volume $V$ is \cite{EarlyUniverse}
\begin{equation}
T dS = d(\rho V)+p dV - \mu d(nV),
\end{equation}
where $\rho$ is the energy density, $p$ is the pressure and $n$ is the number density. The first two terms on the right hand side vanishes according to the total energy conservation.  For the last term, there could be contributions from particle species with non-zero chemical potential, but $d(nV)=0$ for neutrinos because it has been already frozen out and $d(nV)$ is negligible for antineutrinos because its number density is exponentially suppressed ($n_{\bar{\nu}}=(T^3/\pi^2)e^{\xi}$ for $\xi\ll -1$) for such large degeneracy as to make the neutrino number freeze out before the muon-antimuon annihilation ends.

The total entropy conservation is expressed as
\begin{equation}
a_{f}^3 s_{tot} \left( T_{f}(\xi_{\rm initial}), \xi_{\rm initial} \right)=a_{\rm final}^3 s_{tot}(T_{\rm final} ,\xi_{\rm final}), \label{eq:entropyconservation}
\end{equation}
where the subscript "$f$" denotes the value at the neutrino number freeze-out and "final" after the $\mu^{\pm}$ annihilation but before the $e^{\pm}$ annihilation. $\xi$ is conserved when the chemical equilibrium holds so we have written $\xi_{\rm initial}$ instead of $\xi_f$. $s_{tot}$ is the sum of entropy density $s=(\rho+p-\mu n)/T$ of all the particle species exist. On the right hand side, there exists only the relativistic particles, photons, $e^{\pm}$, and neutrinos one type with degeneracy and 2 types without, so the $T^3$ can be scaled out as
\begin{equation}
s_{tot}(T,\xi) \approx \frac{2\pi^2}{45} T^3 \left( g_{\ast}(\xi) -\xi \frac{45}{4\pi^4}\int \frac{x^2}{e^{x-\xi}+1} dx \right) \equiv \frac{2\pi^2}{45}K(\xi)T^3, \label{eq:stotal1}
\end{equation}
where $g_{\ast}(\xi)$ is the effective degrees of freedom same as the one appeared in Eq.~(\ref{eq:gstar}). We neglect exponentially suppressed antineutrino contribution. When we calculate the left hand side of Eq.~(\ref{eq:entropyconservation}), finite masses of muons ($m_{\mu}=106$ MeV) and pions ($m_{\pi}=135$ MeV) are included in order to treat the annihilation of these particles continuously. We assume the quark-hadron phase transition to occur instantaneously at $T_{\rm QCD}=200$ MeV and the quark-gluon phase contains u, d and s quarks which are well approximated as massless. Then
\begin{eqnarray}
s_{tot}(T,\xi) &\approx& \frac{2\pi^2}{45} T^3 \Bigg( g_{\ast}(\xi)-\xi \frac{45}{4\pi^4}\int \frac{x^2}{e^{x-\xi}+1} dx +\frac{95}{2}\theta(T-T_{\rm QCD})\nonumber \\
& &+4 I^{+}_{\mu}(T)+3I^{-}_{\pi}(T) \theta(T_{\rm QCD}-T)\Bigg) \equiv \frac{2\pi^2}{45}J(T,\xi)T^3, \label{eq:stotal2}
\end{eqnarray}
where $\theta(x)$ is 1 for $x>0$ and  is 0 otherwise. $I^{\pm}_i$ denotes the contribution from the massive particles $i$ as
\begin{eqnarray}
I^{\pm}_i(T)=\frac{45}{4\pi^4} \left[ \int_0^{\infty}dx \frac{x^2}{\exp \left(\sqrt{x^2+\alpha_i^2} \right)\pm 1} \left( \sqrt{x^2+\alpha_i^2}+\frac{x^2}{3\sqrt{x^2+\alpha_i^2}} \right) \right], 
\end{eqnarray}
where $\alpha_i=m_i/T$ and the sign is $+/-$ when $i$ is a fermion/boson.

\begin{figure}
\includegraphics{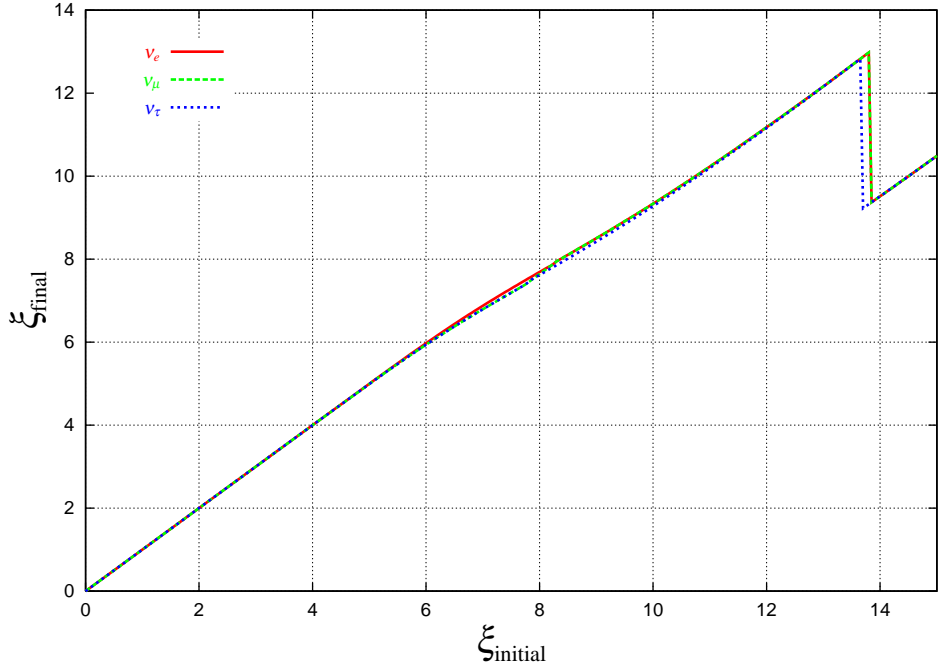}
 \caption{The relation between the degeneracy which exists initially ($\xi_{\rm initial}$) and which remains after the muon-antimuon annihilation ($\xi_{\rm final}$).  For $\xi_{\rm initial}\gtrsim 6$, since the neutrino number freezes while the universe contains relativistic degrees of freedom of the particles other than photons, $e^{\pm}$ and neutrinos, $\xi_{\rm final}$ becomes smaller than $\xi_{\rm initial}$.}
\label{fig:xijump}
\end{figure}

The neutrino number conservation is
\begin{equation}
a_f^3 n_{\nu} \left( T_{f}(\xi_{\rm initial}), \xi_{\rm initial} \right)=a_{\rm final}^3 n_{\nu}(T_{\rm final},\xi_{\rm final}), \label{eq:numberconservation}
\end{equation}
and $n_{\nu}$ is calculated by Eq.~(\ref{eq:nnu}). Dividing each sides of Eq.~(\ref{eq:entropyconservation}) by Eq.~(\ref{eq:numberconservation}), we obtain an equation with respect to $\xi_{\rm final}$ for given $\xi_{\rm initial}$,
\begin{equation}
J(T_f(\xi_{\rm initial}),\xi_{\rm initial})M(\xi_{\rm final})-K(\xi_{\rm final})M(\xi_{\rm initial})=0.
\end{equation}
We numerically solve this for $\xi_{\rm final}$ in the range $0\le\xi_{\rm initial}\le15$. The result is shown in Fig.~\ref{fig:xijump} and we see considerable difference between the initial $\xi_{\nu}$ and the final $\xi_{\nu}$. 

\begin{figure}
\includegraphics{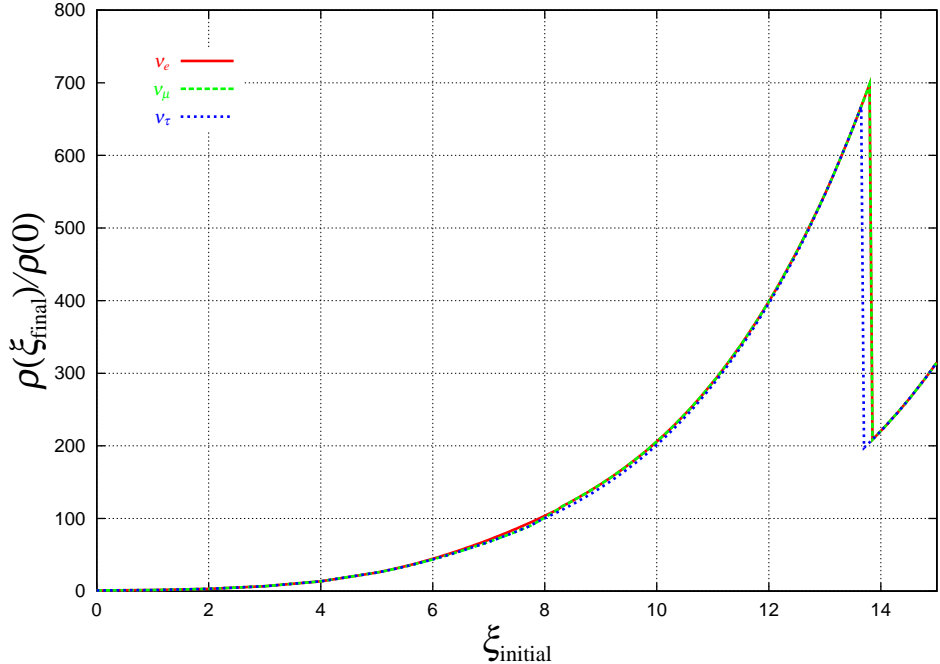}
\caption{The final energy density of the degenerate type of neutrinos and antineutrinos normalized to the one with no degeneracy: $\rho_{\nu+\bar{\nu}}(\xi)/\rho_{\nu+\bar{\nu}}(0)$.}
\label{fig:rhojump}
\end{figure}

 We translate this results in terms of neutrino energy density as shown in Fig.~\ref{fig:rhojump} (with regard to only the degenerate family). We show the final value of the energy density of the degenerate neutrinos plus antineutrinos divided by the one with no degeneracy (so-called effective number of neutrino types minus two).

\begin{figure}
\includegraphics{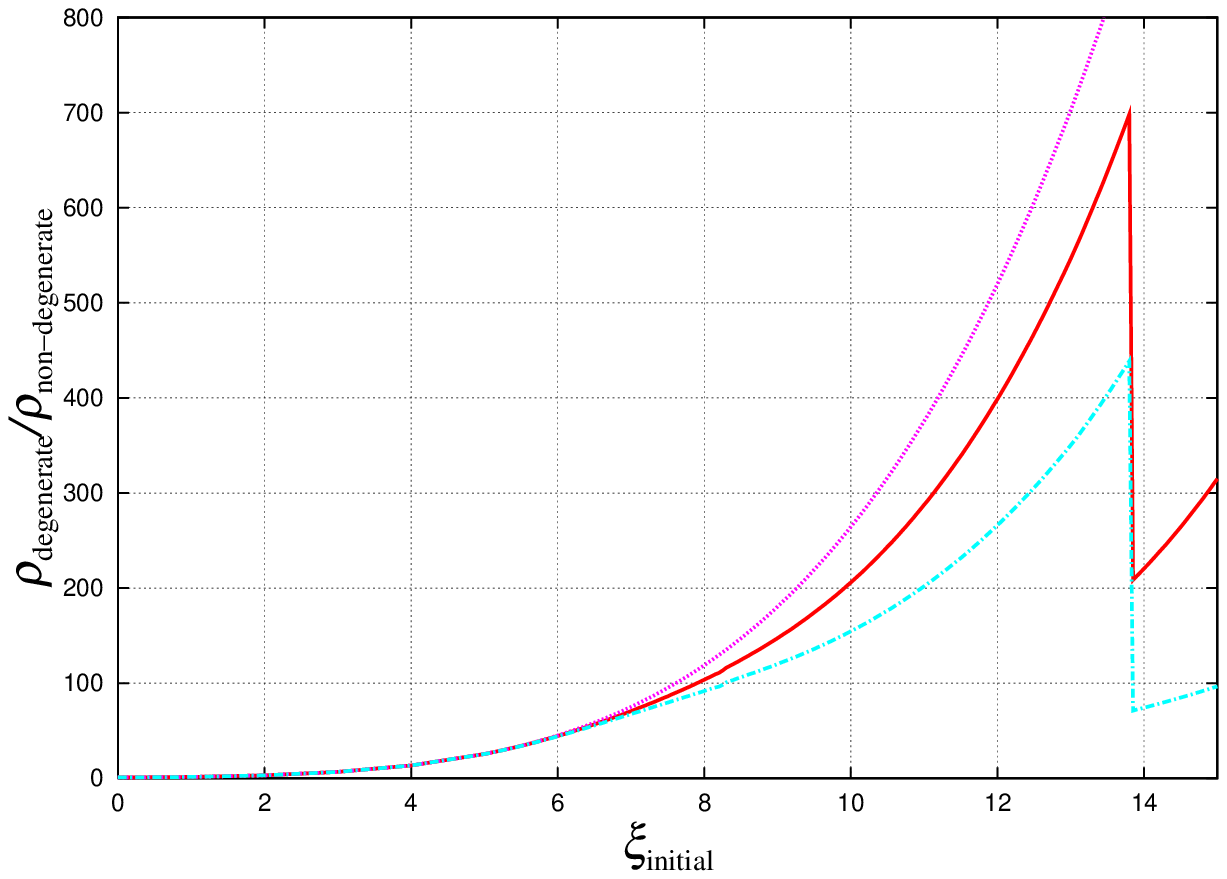}
\caption{A comparison with the results which appear under different assumptions (for $\nu_{e}$). The solid line is our result. The dotted line is the energy density calculated from the initial degeneracy. The dot-dashed line is the one expected when the neutrinos are assumed to be kinematically decoupled simultaneously with its number freeze-out.}
\label{fig:rhojumpcomp}
\end{figure}

For a comparison, in Fig.~\ref{fig:rhojumpcomp}, we also show the degenerate neutrino final energy densities when some assumptions are different. We compute the case when the degeneracy parameter $\xi$ maintains the initial value and the case when we assume, as in Refs.~\cite{Kang1992}\cite{Orito2000}, the neutrinos are not  heated by annihilation processes after its number has frozen out. Our result demonstrates that for some range of the degeneracy, the final neutrino energy density can be lower than the one with the initial degeneracy but not as low as previous analysis \cite{Orito2000} has shown.

At last, we make a comment on the evolution of the antineutrino degeneracy parameter, $\xi_{\bar{\nu}}$. At the initial stage, the chemical equilibrium holds so $\xi_{\bar{\nu}}=-\xi_{\nu}$. This relation could be broken after the neutrino number freeze-out because the neutrino is not in the chemical equilibrium any more. However, since the frequent elastic scattering with electrons ensures neutrinos and antineutrinos to be in kinematical equilibrium and annihilation is still efficient for antineutrinos, the relation $\xi_{\bar{\nu}}=-\xi_{\nu}$ continues to hold. Therefore, as $\xi_{\nu}$ decreases during for example the muon-antimuon annihilation, $\xi_{\bar{\nu}}$ follows its variation with the sign opposite. Meanwhile, when the temperature drops down to the electron-positron annihilation, the elastic scattering is not efficient to keep $\nu$ (and also $\bar{\nu}$ if the degeneracy is not very large) in the kinetic equilibrium with $e^{\pm}$. In this case, the momentum distribution is distorted from the equilibrium form so, strictly speaking, the notion of the chemical potential or the degeneracy parameter is lost. But the distortion can usefully expressed as momentum dependent chemical potential and in this sense, $\xi_{\nu}+\xi_{\bar{\nu}}\neq 0$ as shown in Ref.\cite{Esposito2000} where the evolution of the neutrino spectrum is fully simulated with $\xi \lesssim 1$.  

\section{conclusion}\label{sec:conclusion}

We have re-investigated some properties of the thermal history of the early universe with very large neutrino degeneracy. We have justified and adopted the correct  picture of the neutrino number freeze-out as explained by Ref.~\cite{Dolgov2002}. We have made some complementary arguments and calculations to those found in Ref.~\cite{Dolgov2002} and have obtained the results concerning the evolution of the strong neutrino degeneracy and in turn its energy density. We find there are cases that the neutrino degeneracy parameter $\xi$ becomes smaller than the initial value after the muon annihilation. However, they do not seem to require alteration to the cosmological bounds on the neutrino degeneracy such as $-0.01\leqslant \xi_{\nu_e}\leqslant 0.22$ and $|\xi_{\nu_{\mu,\tau}}|\leqslant 2.6$ \cite{Hansen2001} (more stringent bound, $|\xi_{\nu}|\lesssim 0.07$ for all three neutrino types, is likely to apply taking into account neutrino oscillations with maximal mixing \cite{Dolgov2002a}. The analysis on how the bounds are modified for the region of the Large Mixing Angle solution to solar neutrino problems is found in Refs.~\cite{Wong2002} and \cite{Abazajian2002}), because, in Fig.~\ref{fig:xijump}, the local minimum of the final degeneracy caused by the quark-hadron phase transition does not have the value lower than the present upper bound. 

Finally, we note our results stem from the fact that the degenerate neutrinos are in kinetic equilibrium with the rest of the cosmic plasma well after their number has frozen out. The same is true for other possibly degenerate species of stable particles other than the neutrinos and our analysis is applicable to them. So, although our analysis on the neutrinos has turned out not to affect the cosmological bound on their degeneracy, the remark made in this paper is worth bearing in mind.


\begin{thebibliography}{}
\bibitem{Dolgov2002} A. D. Dolgov, Phys. Rep. {\bf 370}, 333 (2002).
\bibitem{Kang1992} H-S. Kang and G. Steigman, Nucl. Phys. {\bf B372}, 494 (1992).
\bibitem{Orito2000} M. Orito, T. Kajino, G. J. Mathews, and R. N. Boyd, astro-ph/0005446.
\bibitem{Freese1983} K. Freese, E. W. Kolb, and M. S. Turner, Phys. Rev. D {\bf 27},1689 (1983).
\bibitem{EarlyUniverse} E. W. Kolb and M. S. Turner, {\it The Early Universe} (Addison Wesley, Reading, MA, 1990).
\bibitem{Esposito2000} S. Esposito, G. Miele, S. Pastor, M. Peloso and O. Pisanti, Nucl. Phys. {\bf B590}, 539 (2000).
\bibitem{Orito2002} M. Orito, T. Kajino, G. J. Mathews, and Y. Wang, Phys. Rev. D {\bf 65}, 123504 (2002).
\bibitem{etc} J. Lesgourgues and S. Pastor, Phys. Rev. D {\bf 60}, 103521 (1999);
                        S. Pastor and J. Lesgourgues, Nucl. Phys. Proc. Suppl. {\bf 81}, 47 (2000).
\bibitem{Hansen2001} S. H. Hansen, G. Mangano, A. Melchiorri, G. Miele, and O. Pisanti, Phys. Rev. D {\bf 65}, 023511 (2001).
\bibitem{Dolgov2002a} A. D. Dolgov, S. H. Hansen, S. Pastor, S. T. Petcov, G. G. Raffelt, and D. V. Semikoz, Nucl. Phys. {\bf B632}, 363 (2002). 
\bibitem{Wong2002} Y. Y. Y. Wong, Phys. Rev. D {\bf 66}, 025015 (2002).
\bibitem{Abazajian2002} K. N. Abazajian, J. F. Beacom, and N. F. Bell, Phys. Rev. D {\bf 66}, 013008 (2002).
\end{thebibliography}
 \end{document}